# The role of the "Maximizing Output Growth Inflation Rate" in monetary policy

Dominique R.D. Pépin[*]

I.R.I.A.F., Niort, France and C.R.I.E.F., Poitiers, France

**Abstract**

The paper discusses the role of monetary policy when potential output depends on the inflation rate. If the intention of the central bank is to maximize actual output growth, then it has to be credibly committed to a strict inflation targeting rule, and to take the MOGIR (the Maximizing Output Growth Inflation Rate) as the target.

## 1. Introduction

What is a "good monetary policy"? The answer to this question seems to achieve unanimity among monetary economists: « "good monetary policy" is used in the conventional meaning of successfully stabilizing inflation around a low average level, with some concern for stabilizing output around potential output » (Svensson, 2003, p. 426). According to a commonly accepted point of view, the goal of the central banker is to stabilize both inflation and output[1]. Such an objective is however defined in a more or less arbitrary way[2], and can be disputed. Some authors suggest rationalizing the behavior of the central banker, by formulating the goal of monetary policy from the utility of a representative agent (e.g. Aiyagari and Braun, 1997, Ireland 1996 and Rotemberg and Woodford, 1997), but this approach is of a limited interest. On the one hand, these models do not take into account the major cost of inflation, the uncertainty which its variability generates; and on the other hand, the utility of a representative agent constitutes a bad guide to welfare analysis when there are incomplete insurance and credit markets (Clarida, Gali and Gertler, 1999). Is it then really rational to base monetary policy on an objective of pure stabilization? Why does the central bank have to attach so much importance to the fight against the inflation, often defined as the priority objective of the stabilization policy, or even as the ultimate objective of the central banker?

---

[*] Tel. : +33-5-49-28-75-51 ; fax : +33-5-49-28-14-49
  E-mail address: Dominique.Pepin@mshs.univ-poitiers.fr (D.R.D Pépin)
  1. The possibility that the central banker can pursue personal or different objectives from those of the society is omitted here.
  2. See Rotemberg and Woodford (1999) and Woodford (1999) for a justification of this kind of policy.

According to Di Tella, MacCulloch and Oswald (2001), the central bank gives an exaggerated importance to inflation compared to unemployment, if we judge it by the effects of inflation and unemployment on the happiness of the agents. Unemployment exerts a powerful effect on satisfaction of the agents: the higher unemployment is, the more the agents fear to lose their jobs, and the more they suffer with the idea that a large number of their fellow-citizens do not succeed in finding employment. Inflation also exerts a significant effect on the agent's happiness, but less than that of unemployment. How can we reconcile these facts with the top priority given by economists, governments and central bankers, the fight against inflation?

There are two answers to this question: either the position taken by economists is irrational, or mechanisms exist which are absent from their models but which are taken into account in an indirect way, and which confer on inflation a significant influence on activity and employment. It is this second answer which we prefer.

Indeed, the growth effects of inflation justify the objective of inflation stabilization. In a traditional backward-looking keynesian model, we add a relation which sensitizes potential output growth to the difference of the actual inflation rate with a critical inflation rate called MOGIR (for Maximizing Output Growth Inflation Rate). The more the actual inflation rate deviates from the MOGIR, the weaker potential output growth is. If the MOGIR is strictly positive and not too high, that means that a bit of inflation can contribute to growth. If it is null, then inflation always harms the growth.

A weakness of our analysis is the ad hoc character of this relation between inflation and growth, to which we do not give any rigorous microeconomic foundation. In spite of this absence, it is important to analyze the consequences of this relation between growth and inflation, because in our opinion it is in conformity with the beliefs of many economists, politicians, and especially of central bankers, who have to make daily monetary policy decisions.

The introduction of this relation between inflation and growth rationalizes the fight against inflation. It makes it possible to redefine the objective of monetary policy, the fight against inflation being brought back to the legitimate row of means, and not as an end in itself. We show that it is optimal for the central bank to target the MOGIR. We find the result of Svensson (1997, 1999, 2002, 2003), obtained within the framework of more conventional models, according to which inflation forecast targeting is optimal, but with however the difference that it is a strict inflation target which must be defined, and not a flexible inflation target.

Section II presents the model and its resolution under the traditional objective of pure stabilization of inflation and output gap. Section III analyzes monetary policy when the objective is to maximize actual output growth. Section IV shows that the central banker may find it beneficial to commit to a strict inflation targeting rule. Section V compares the various strategies of monetary policy, and section VI concludes.

**2. Presentation of the model and resolution under an objective of pure stabilization of inflation and output gap**

The model that we use is a very simple backward-looking model, to avoid any useless complication, and because this kind of model (in opposition to forward-looking models) doubtless corresponds to the pragmatic monetary policy knowledge of central bankers (Blinder, 1997)[3].

Consider therefore the model:

$$x_t = \beta x_{t-1} - \phi(i_{t-1} - E_{t-1}\pi_t - r) + \varepsilon_{xt}, \tag{1}$$

$$\pi_t = \pi_{t-1} + \lambda x_t + \varepsilon_{\pi t}, \tag{2}$$

$$x_t = y_t - y_t^p, \tag{3}$$

where $x_t$ is the output gap, the difference between actual output $y_t$ and the potential output level $y_t^p$ (the level of output that would result with flexible prices), $i_{t-1}$ is the monetary policy instrument (a short-term interest rate fixed by the central bank), $\pi_t = p_t - p_{t-1}$ is the inflation rate in period t, and $p_t$ is the log price level. $E_{t-1}(.)$ denotes expectations conditional upon information available[4] at the end of the period $t-1$. $\varepsilon_{xt}$ and $\varepsilon_{\pi t}$ are i.i.d. shocks in period $t$ that are not known in period $t-1$, with constant conditional variances: $E_{t-1}\varepsilon_{xt} = E_{t-1}\varepsilon_{\pi t} = 0$, $V_{t-1}\varepsilon_{xt} = \sigma_x^2$, $V_{t-1}\varepsilon_{\pi t} = \sigma_\pi^2$ $\forall t$, and they are uncorrelated: $E_{t-1}(\varepsilon_{xt}\varepsilon_{\pi t}) = 0$.

---

3 . Backward-looking models can in addition be preferred with forward-looking models for their best empirical fit (Estrella and Fuhrer, 2002, 2003).

The model made up of equations (1) to (3) is a completely conventional monetary model[5], to which we add the additional equation:

$$y_t^p = \delta - \gamma E_{t-1}(\pi_t - \pi^n)^2 + y_{t-1}^p + \varepsilon_{yt}, \qquad (4)$$

where $\varepsilon_{yt}$ is i.i.d.: $E_{t-1}\varepsilon_{yt} = 0, V_{t-1}\varepsilon_{yt} = \sigma_\varepsilon^2$ $\forall t$, and uncorrelated with the other shocks: $E_{t-1}(\varepsilon_{yt}\varepsilon_{xt}) = E_{t-1}(\varepsilon_{yt}\varepsilon_{\pi t}) = 0$ $\forall t$. All the parameters $\beta, \phi, \lambda, \delta, \gamma, r$ of the model are strictly positive, and $\lambda$ and $\phi$ in addition fulfill $\lambda < 1, \phi < 1$. For our purposes, it is supposed that these parameters are given. To facilitate later calculations, it is useful to solve both equations (1) and (2) with regard to $x_t$ and $\pi_t$:

$$x_t = \frac{\beta x_{t-1} - \phi(i_{t-1} - \pi_{t-1} - r)}{1 - \lambda\phi} + \varepsilon_{xt}, \qquad (5)$$

$$\pi_t = \frac{\pi_{t-1} + \lambda[\beta x_{t-1} - \phi(i_{t-1} - r)]}{1 - \lambda\phi} + \lambda\varepsilon_{xt} + \varepsilon_{\pi t}. \qquad (6)$$

According to equation (4), the potential output growth $y_t^p - y_{t-1}^p$ depends on the difference of the expected inflation rate with the critical inflation rate $\pi^n$, called MOGIR (Maximizing Output Growth Inflation Rate)[6]. The MOGIR is the inflation rate which maximizes potential output growth. Indeed, if $\pi_t = \pi^n$ $\forall t$, potential output growth takes the maximum value $y_t^p - y_{t-1}^p = \delta + \varepsilon_{yt}$. And any difference between the actual inflation rate and the MOGIR reduces potential output growth.

---

4. It is supposed that this information is the same for the central banker and the private sector, and that anticipations are rational.

5. See Fuhrer (1994, 1997), Fuhrer and Moore (1995a, 1995b), Svensson (1997) and Rudebusch and Svensson (2002) for presentations of backward-looking models.

6. Equation (4) is to be compared with that of Smyth (1992): $y_t^p - y_{t-1}^p = \delta - \gamma\pi_t + \varepsilon_{yt}$, where inflation is a nuisance for growth.

Many economists and political decision-makers share the feeling that a certain amount of inflation is necessary to promote a sustained high growth. Admittedly, this idea is highly debated[7]. According to some authors, inflation always harms growth, however weak it is. Kormendie and Meguire (1985), Grier and Tullock (1989), Fischer (1991) and Smyth (1994) bring empirical evidence confirming this thesis. Potential output growth depends negatively on inflation according to Smyth (1992) and Schaling and Smyth (1994). For other authors, moderate inflation rates have a positive impact on growth. Bullard and Keating (1995) show that if inflation exerts a negative effect in countries with strong inflation, it exerts on the contrary a positive effect in countries with weak inflation. All the authors however agree to recognize that deflation has an adverse impact on growth. To synthesize the thought of all these authors, we can suppose that the relation between inflation and growth is not linear. When the actual inflation rate is higher than the critical inflation rate $\pi^n$, then inflation is a nuisance for growth; and on the contrary when it is lower than $\pi^n$, it contributes to growth. So, these authors differ by the value which they attribute to $\pi^n$ (the MOGIR). When $\pi^n = 0$, the growth effects of inflation are negative; they are positive when $\pi^n > 0$.

At first, let us consider the traditional behavior of pure stabilization of the central banker, acting so as to stabilize output gap and inflation. We suppose that the central banker thinks in a short time frame, what is doubtless rather close to reality (Blinder, 1997). Following much of the literature, we assume that the policy problem is to minimize the following loss function:

$$L(x_t, \pi_t) = E_{t-1} x_t^2 + \alpha E_{t-1} (\pi_t - \pi^n)^2 , \qquad (7)$$

where $\alpha > 0$ is the relative weight on output stabilization.

According to equation (7), the MOGIR represents the long-run inflation target, and the long-run output gap target is null[8].

The first-order condition for minimizing equation (7) over $i_{t-1}$ can be written:

---

7. See Temple (2000) for a review of the literature on growth and inflation.

8. This choice is coherent with the idea that actual output cannot deviate in the long run from potential output, because monetary policy has no long-term effect on output.

$$E_{t-1}x_t + \alpha\lambda E_{t-1}(\pi_t - \pi^n) = 0. \tag{8}$$

We take the conditional expectations of equations (5) and (6), that we incorporate in equation (8), to obtain the optimal reaction function of the central bank:

$$i_{t-1} = r + \pi^n + \frac{\beta}{\phi}x_{t-1} + \frac{(\phi + \alpha\lambda)}{\phi(1+\alpha\lambda^2)}(\pi_{t-1} - \pi^n). \tag{9}$$

This reaction function is of the same form as the Taylor rule (Taylor, 1993). The monetary policy instrument $i_{t-1}$ is increasing in the excess of actual inflation over $\pi^n$, and in actual output over the potential output level. In addition this feedback rule satisfies the Taylor principle[9], since according to the parametric assumptions, the coefficient on $\pi_{t-1} - \pi^n$ exceeds unity: $(\phi + \alpha\lambda)/\phi(1+\alpha\lambda^2) > 1$. Then using (9) to eliminate $i_{t-1}$ in the reduced form, we get the output gap and inflation equilibrium values:

$$x_t = -\frac{\alpha\lambda}{(1+\alpha\lambda^2)}(\pi_{t-1} - \pi^n) + \varepsilon_{xt}, \tag{10}$$

$$\pi_t = \pi_{t-1} - \frac{\alpha\lambda^2}{(1+\alpha\lambda^2)}(\pi_{t-1} - \pi^n) + \lambda\varepsilon_{xt} + \varepsilon_{\pi t}. \tag{11}$$

By adjusting interest rates according to equation (9), the central bank is able to hit its long-run inflation and output targets simultaneously, all the time: $Ex_t = 0, E\pi_t = \pi^n$. These results, which neglect the contribution of equation (4), are standard results widely described in monetary economy literature.

For later comparisons, we calculate expected output growth:

$$E_{t-1}(y_t - y_{t-1}) = \delta - \gamma\left[\left(\frac{\pi_{t-1} - \pi^n}{1+\alpha\lambda^2}\right)^2 + \lambda^2\sigma_x^2 + \sigma_\pi^2\right] - \frac{\alpha\lambda}{(1+\alpha\lambda^2)}(\pi_{t-1} - \pi^n) - x_{t-1}, \tag{12}$$

and long-run output growth:

$$E(y_t - y_{t-1}) = \delta - \gamma Var\pi_t, \quad Var\pi_t = \frac{(1+\alpha\lambda^2)^2(\lambda^2\sigma_x^2 + \sigma_\pi^2)}{(1+\alpha\lambda^2)^2 - 1}. \tag{13}$$

---

9. An interest-rate rule satisfies the Taylor principle (Taylor, 1999) if it implies that, in the event of a sustained increase in the inflation rate by k %, the nominal interest rate will be raised by more than k %. It is a criterion for sound monetary policy according to Taylor (1999).

## 3. Maximizing actual output growth

By minimizing the loss function (7), the policy-maker worries only about stabilizing inflation and output. It is a reasonably good objective, but it is completely arbitrary to think that this objective drives the policy-making of every central banker. Indeed, most of the statements or mandates defining the role of central banks reveal, with the objective of inflation stabilization (the target inflation is always defined strictly positive), an objective of promotion of output growth. If this prospect is adopted, the loss function (7) is inappropriate to describe the objectives of central bankers. We suppose in this section that monetary policy aims to maximize $E_{t-1} y_t$, which means maximizing $E_{t-1}(y_t - y_{t-1})$, the expected output growth. The objective of monetary policy is thus clearly an objective of promotion of output growth. The fight against inflation is not then any more an objective in itself. Its effect on growth according to equation (4) incites one however to look for its stabilization. The fight against the inflation is not thus any more a social objective, but the means to favor output growth.

The first-order condition to the maximization problem can be written:

$$E_{t-1} \pi_t = \pi^n + \frac{1}{2\gamma\lambda}. \tag{14}$$

This results in:

$$i_{t-1} = r + \pi^n - \frac{1-\lambda\phi}{2\gamma\lambda^2\phi} + \frac{\beta}{\phi} x_{t-1} + \frac{1}{\lambda\phi}(\pi_{t-1} - \pi^n). \tag{15}$$

We again obtain a reaction function which is of the same type as the Taylor rule, and which satisfies the Taylor principle, because $1/\lambda\phi > 1$.

Using (15) to eliminate $i_{t-1}$ in both equations (5) and (6), we get the output gap and inflation equilibrium values:

$$x_t = \frac{1}{2\gamma\lambda^2} - \frac{1}{\lambda}(\pi_{t-1} - \pi^n) + \varepsilon_{xt}, \tag{16}$$

$$\pi_t = \pi^n + \frac{1}{2\gamma\lambda} + \lambda\varepsilon_{xt} + \varepsilon_{\pi t}. \tag{17}$$

We observe according to equation (17) that long-run equilibrium inflation is forced systematically above the MOGIR: $E\pi_t = \pi^n + 1/(2\gamma\lambda)$. And long-run equilibrium output gap always equals its target:

$Ex_t = 0$. The search for maximal output growth therefore leads to target the inflation rate at a level higher than the MOGIR. In the long run, inflation is thus higher with this strategy than with that of stabilization. We would be tempted to say that this higher inflation is the price to be paid to maximize expected output growth, but in fact it does not truly constitute a cost, because inflation does not enter the objective function of the central banker. This higher inflation thus does not involve a welfare loss for the private sector, and cannot thus be interpreted as an inconvenience of this monetary policy strategy.

Expected output growth will fulfill:

$$E_{t-1}(y_t - y_{t-1}) = \delta - \gamma(\lambda^2 \sigma_x^2 + \sigma_\pi^2) + \frac{1}{4\gamma\lambda^2} - \frac{1}{\lambda}(\pi_{t-1} - \pi^n) - x_{t-1}. \tag{18}$$

According to the definition of the adopted strategy, equation (18) gives us the maximum value of expected output growth. Finally, we deduce that long-run output growth is:

$$E(y_t - y_{t-1}) = \delta - \gamma\left[\left(\frac{1}{2\gamma\lambda}\right)^2 + \lambda^2 \sigma_x^2 + \sigma_\pi^2\right], \tag{19}$$

which on the other hand is not inevitably maximal.

## 4. Inflation forecast targeting

Rather than to maximize actual output growth, the central banker can prefer to maximize potential output growth. Without making any calculation, we observe according to (4) that maximizing $E_{t-1}(y_t^p - y_{t-1}^p)$ means minimizing $E_{t-1}(\pi_t - \pi^n)^2$. Maximizing the expected growth rate of potential output thus amounts to choosing the MOGIR as the inflation target, so to adopting a strict inflation targeting rule[10].

Inflation targeting is now widely described in monetary economy literature[11]. A first advantage of this strategy is that, « in terms of communication, the announcement of inflation target clarifies the central bank's intentions for the markets and for the general public, reducing uncertainty about the future course of inflation » (Bernanke et Mishkin, 1997, p. 106). It increases the transparency and the coherence of monetary policy. It in addition provides the central bank with an implicit mechanism of commitment, while leaving it the possibility of acting discretionarily. A second advantage of this strategy, demonstrated in this section and the following one, is that such a policy leads to higher long-run actual output growth. Finally, inflation targeting has been introduced in recent years in number of industrialized countries. This vision of monetary policy thus seems to more or less correctly describe the real behaviour of central bankers.

---

10. We thus refer here to strict inflation targeting and not to flexible inflation targeting, according to Svensson's terminology (1997).

11. See Bernanke and Mishkin (1997) and Svensson (1997, 1999, 2002), among others, for discussion of inflation targeting.

The first order condition of the minimization problem is:

$$E_{t-1}\pi_t = \pi^n. \tag{20}$$

This condition justifies the expression of inflation forecast targeting, since the instrument should hence be set so as to make the inflation forecast equal to the target. This leads to the optimal reaction function:

$$i_{t-1} = \pi^n + r + \frac{\beta}{\phi}x_{t-1} + \frac{1}{\lambda\phi}(\pi_{t-1} - \pi^n). \tag{21}$$

This rule is very similar with that maximizing expected output growth. It turns out that we obtain equation (21) by simply adding the positive constant $(1-\lambda\phi)/(2\gamma\lambda^2\phi)$ to rule (15). Monetary policy rule (21), compared to the preceding one, thus results fixing the interest rate at a more significant level. The parameters of the optimal reaction function in response to inflationary and output gap are however the same ones for both types of policy.

We then calculate the output gap and inflation rate equilibrium values:

$$x_t = -\frac{1}{\lambda}(\pi_{t-1} - \pi^n) + \varepsilon_{xt}, \tag{22}$$

$$\pi_t = \pi^n + \lambda\varepsilon_{xt} + \varepsilon_{\pi t}. \tag{23}$$

The long-run inflation rate equals the MOGIR: $E\pi_t = \pi^n$, and long-run output gap is null: $Ex_t = 0$. Both strict inflation targeting and pure stabilisation strategies are similar in successfully stabilizing inflation and output in the long run.

We calculate the expected growth rate of potential output:

$$E_{t-1}(y_t^p - y_{t-1}^p) = \delta - \gamma(\lambda^2 \sigma_x^2 + \sigma_\pi^2), \tag{24}$$

then the expected growth rate of actual output:

$$E_{t-1}(y_t - y_{t-1}) = \delta - \gamma(\lambda^2 \sigma_x^2 + \sigma_\pi^2) - \frac{1}{\lambda}(\pi_{t-1} - \pi^n) - x_{t-1}, \tag{25}$$

and finally the long-term growth rate of actual output:

$$E(y_t - y_{t-1}) = \delta - \gamma Var\pi_t, Var\pi_t = \lambda^2 \sigma_x^2 + \sigma_\pi^2. \tag{26}$$

## 5. Comparison of the various strategies of monetary policy

We now will compare the various strategies of pure stabilization, maximization of expected output growth, and strict inflation targeting. We start by comparing the capacity of these strategies to stabilize the long-term output gap and inflation. Table 1 recapitulates the results obtained with the various strategies of monetary policy.

These policies differ in the induced long-run inflation level. The strategies of pure stabilization and strict inflation targeting are from this point of view identical, because both imply that the long-run inflation rate is equal to the MOGIR. So both are identical in successfully stabilizing inflation and output. On the other hand, maximizing expected output growth led to an inflation rate higher than the MOGIR[12]. This policy is more inflationary in the long run.

We compare then the various policies in term of expected output growth (table 2). To maximize expected output growth, it is obviously the strategy conceived for this purpose which is the best. The two other strategies, of pure stabilization and strict inflation targeting, lead to different and inevitably worse results than the first. If the ultimate objective of monetary policy is to support output growth, then the central banker should rather maximize $E_{t-1}(y_t - y_{t-1})$.

Finally, what do these strategies imply for long-term output growth (table 3)?

According to the results of table 3, it appears that the highest long-run output is the one which ensues from strict inflation targeting, and not from maximizing expected output growth. So, in a long term perspective, the central banker should rather adopt a strict inflation targeting rule. However, every period, he may find it beneficial to maximize expected output growth according to results of table 2. The strict inflation targeting rule is thus not time consistent, while at the same time it is the best possible long-run strategy if the ultimate objective of monetary policy indeed is to promote output growth. It explains the need for the legislator to register this objective of strict inflation targeting in the central banker's mandate, so that he can commit himself credibly to such a rule[13].

---

12. Let us remember that this result does not constitute in itself a condemnation of this policy, because inflation does not appear in the central banker's objective function.

## 6. Conclusion

Is a good monetary policy only a policy which allows for the stabilization of inflation and output in an optimal way? We answered by the negative in this paper, explaining that the ultimate goal of any policy is to maximize social welfare. If we admit that social welfare increases with activity and production (and correlatively with employment), then the ultimate objective of monetary policy must be to promote output growth. Does this mean that the central bankers must ignore inflation? Certainly not, if we admit that inflation affects potential output growth. The fight against inflation is then always in the heart of monetary policy, but this time as a means to satisfy the stated objectives, and not as the objective in itself.

A good monetary policy is then a policy of promotion of output growth, and in this perspective, the central banker should rather commit himself credibly to a strict inflation targeting rule. The inflation target must be the MOGIR, the one which maximizes output growth.

Strict inflation targeting makes it possible on the one hand to stabilize long-run inflation and output gap, as a policy of pure stabilization would do, and on the other hand to maximize output growth. We think that the model presented in this paper is rather close to the monetary intuitions of central bankers, explaining the attraction recently observed for inflation targeting rules.

---

13. In Svensson's model (1997), inflation targeting is the optimal monetary policy; it is time consistent, and "there is no specific need to monitor monetary policy in order to ensure that the central bank implements inflation targeting" (Svensson, 1997, p. 1121). Inflation targeting thus does not need to be inevitably registered in the mandate of the central banker, who will pursue this policy anyway.

Lastly, if we trust the results presented in this paper, optimal monetary policy requires to have a good guess of the MOGIR. Is the MOGIR equal to 0%, 2%, 3%…? It is an empirical question which, in the light of our analysis, is of paramount importance, and to which the researchers in monetary economy should devote significant efforts to try to answer.